\documentclass[12pt]{iopart}
\usepackage{graphicx}

\begin{document}

\title[Baryon to Meson Ratios of Jets at STAR]{Baryon to Meson Ratios on
the Near and Away-Side of Jets and their Centrality Dependence at
STAR}

\author{Jiaxu Zuo (for the STAR Collaboration)}

\address{Nuclear Physics Division, Shanghai Institute of
Applied Physics, CAS.\\
Shanghai 201800, P. R. China\\
}

\ead{zuojiaxu@sinap.ac.cn}

\begin{abstract}
We measure relative abundances of $K_{S}^{0}$, $\Lambda$ and
$\bar{\Lambda}$ in near-side and away-side cones correlated with
triggered high-$p_{T}$ particles in $^{197}$Au + $^{197}$Au
collisions at $\sqrt{s_{NN}}$ = 200 GeV. The centrality dependence
of identified particles in the triggered particle cones is also
presented. Particle yields and ratios are extracted on the
near-side and away-side of the trigger particle. The
associate-particle ratios are studied as a function of the angle
relative to the trigger particle azimuth $\Delta\phi$. Such
studies should help elucidate the origin of the modifications in
the jet like correlations observed in Au+Au collisions relative to
p+p collisions. And these studies also will help understand the
variation of local parton densities at the away side. We discuss
how these measurements might be related to several scenarios for
interactions of fast partons with the medium created in Au+Au
collision.
\end{abstract}

%\keyword{Di-hadron correlation, Mach cone, STAR, RHIC
%Recombination}

\section{Introduction}\label{intro}
The observation of large collective
flows~\cite{STAR_PRL90_flow,STAR_PRL90_flow2} and
jet-quenching~\cite{STAR_PRL90_Hardtke,STAR_PRL91_highpt}
indicates that a dense medium is created in Au+Au collisions at
RHIC~\cite{STAR_NPA_whitepapers}. Studies of two particle
azimuthal correlations have revealed detailed information about
jet interactions with this medium
\cite{STAR_PRL90_Hardtke,STAR_PRC73_2006,STAR_PRC75_2007}. These
measurements can be used to infer properties of the medium such as
temperature, density, and viscosity.

In di-hadron correlations from central Au+Au collisions the
away-side jet opposing the high $p_{T}$ triggered particle
disappears~\cite{STAR_PRL90_Hardtke,STAR_PRL91_highpt}, while the
remnants of the away-side jet are recovered at lower $p_T$
values~\cite{STAR_PRL95_Fuqiang}. The distribution of these
remnants in $\Delta\phi$ is highly modified in comparison to p+p
collisions: the away-side correlation is no longer peaked at
$\Delta\phi=\pi$ but instead has two peaks shifted to either side
of $\pi$~\cite{volcano}. Several scenarios have been proposed to
account for this
splitting~\cite{Mach_Casalderrey-Solana,Mach_Stocker,Deflect_Armesto,cherenkov_Koch,MaGL_AMPT}.
And the analysis of three particles correlations is being pursued
as one method to differentiate scenarios from jet deflection to
cone emission of
particles~\cite{tri-hadron_Ulery,tri-hadron_Ajitanand_PHENIX}.

More information may be obtained about the interaction of fast
partons with the medium by studying the particle-type composition
of the di-hadron correlations. An increase in the ratio of baryons
to mesons has been observed in Au+Au collisions~\cite{P.Sorensen}.
This increase may depend on the parton density of the system. The
coalescence of constituent quarks has been used to describe
successfully much of the observed phenomena. By extension, one
might expect a larger baryon-to-meson ratio for intermediate $p_T$
hadrons on the away-side due to the coalescence of quenched
fragments with each other or with constituents from the medium.
Studies of the $p/\pi$ ratio show evidence for such an
effect~\cite{PHENIX_PRL91_ppbar}.

Information about the relative contribution of quarks and gluons
may also be inferred from the antibaryon-to-baryon ratio: the
fragmentation of gluon jets yields a larger antibaryon-to-baryon
ratio than the fragmentation of quark jets~\cite{quarkvsgluon}.
For this reason, if the splitting of the away-side jet is linked
to large-angle gluon radiation, then the antibaryon-to-baryon
ratio should increase at angles away from $\Delta\phi=\pi$. The
presence of these gluons may also contribute to an increase in the
baryon-to-meson ratio: a recent study found that the baryon
density is largest in collision processes involving gluons (i.e.
qg, gg, q$\overline{\mathrm{q}}$g, or
ggg)~\cite{Zhangbu_BaryonDensity}. For these reasons, measurements
of the baryon-to-meson ratio and the antibaryon-to-baryon ratio on
the near- and away-side of jets should be useful for understanding
the interaction of fast partons with the medium.

We will present measurements of di-hadron correlations of
unidentified trigger hadrons with identified $K_S^0$, $\Lambda$,
or $\overline{\Lambda}$ associated partners. For this analysis, a
trigger hadron is any charged track with $3<p_T<6$~GeV/c while
associated partners are taken from $1<p_T<4$~GeV/c. In the
$|\Delta\eta|<1$ range, we get the yield $dN/d\Delta\phi$ for
$K_{S}^{0}$, $\Lambda$ and $\bar{\Lambda}$ as a function of
$\Delta\phi$. The same procedure is carried out on a mixed event
sample to obtain a background distribution used to correct for
imperfect detector acceptance.

The $v_{2}$ modulated background distribution is subtracted from
the corrected $dN/d\Delta\phi$ distribution by zero-yield at the
minimum (ZYAM)~\cite{PHENIX_PRL97_ZYAM} or zero-yield at
$\Delta\phi=1$ (ZYA1). The following form is used to describe the
$v_2$ modulated combinatorial
background~\cite{STAR_PRL90_flow,PHENIX_PRL97_ZYAM}:
$B(\Delta\phi)=b_{0}(1+2\langle v_{2}^{A}\times
v_{2}^{B}\rangle\cos(2\Delta\phi))$. The nominal $v_2$ is taken as
the average of $v_2$ from an event plane analysis
($v_{2}\{EP\}$)~\cite{STAR_PRL92_partdepend} and $v_2$ from a
4-particle cumulant analysis ($v_{2}\{4\}$)~\cite{STAR_PRC66_v24}
for the charged hadron or $v_{2}$ from the Lee-Yang Zero method
analysis ($v_{2}\{LYZ\}$)~\cite{LYZ} for $K_S^0$, $\Lambda$, or
$\overline{\Lambda}$~\cite{STAR_PRL89_KsLv2,STAR_PRL95_Flow_KL}.
The difference between $v_{2}\{4\}$($v_{2}\{LYZ\}$) and
$v_{2}\{EP\}$ results and the $v_2$
fluctuations~\cite{STAR_v2fluctuation} are considered in the
systematic errors.

\section{Results}\label{result}

\begin{figure}[htb]
\vspace{-3pt}
\resizebox{0.333\textwidth}{!}{\includegraphics{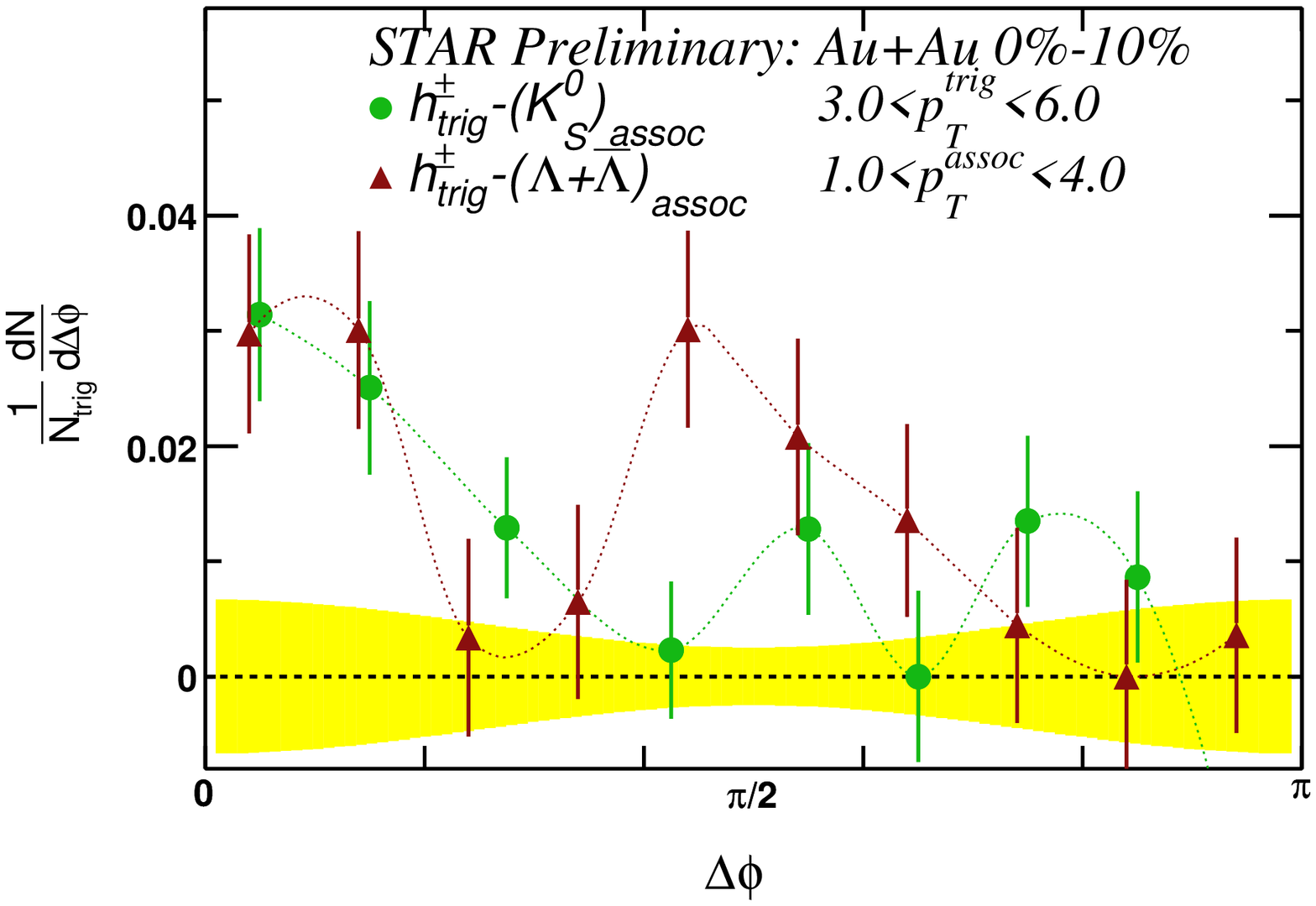}}
\resizebox{0.333\textwidth}{!}{\includegraphics{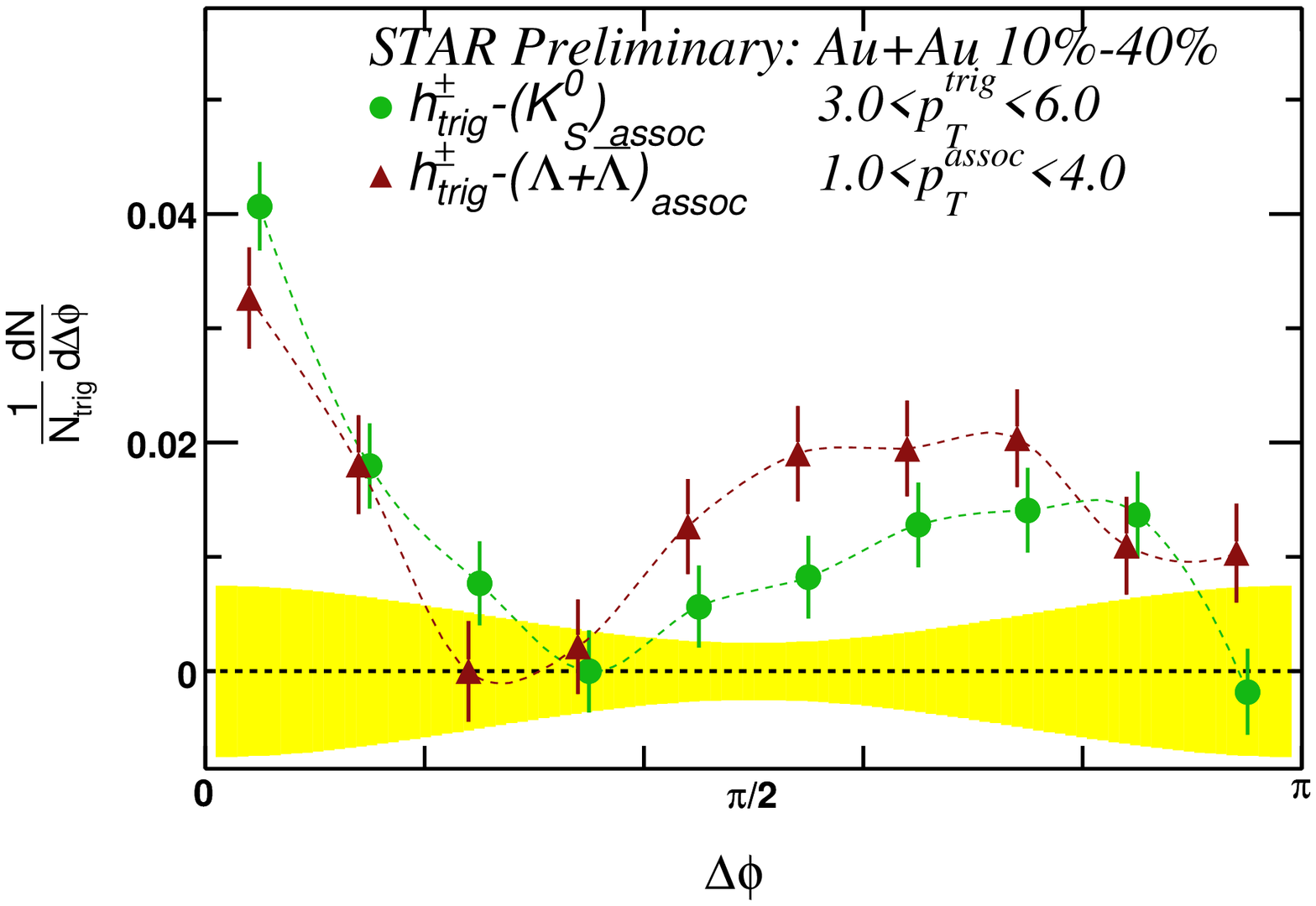}}
\resizebox{0.333\textwidth}{!}{\includegraphics{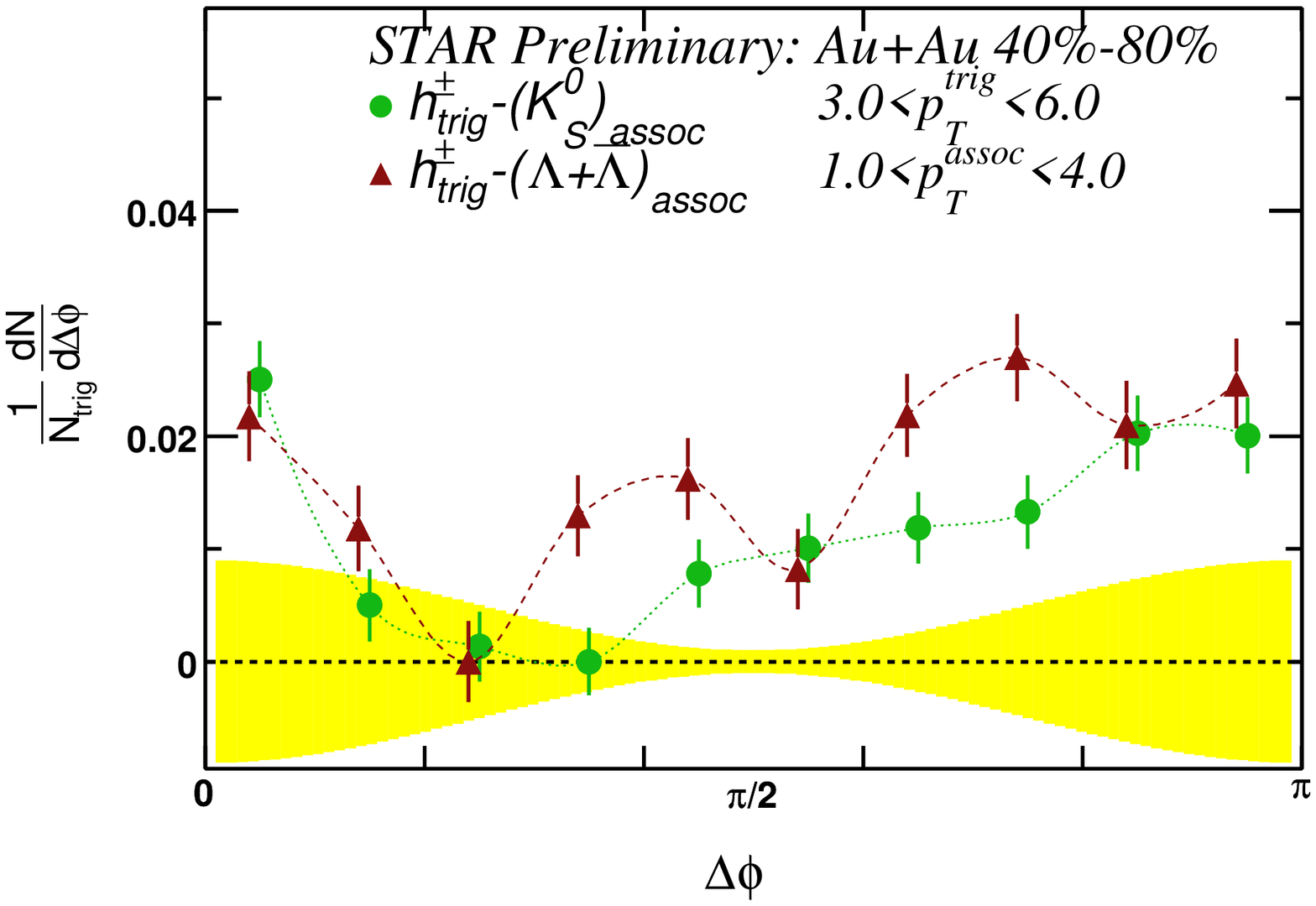}}
\caption[]{Hadron-$K_{S}^{0}$, -$\Lambda$+$\overline{\Lambda}$
correlation function from central to peripheral collision in
$^{197}$Au + $^{197}$Au collisions at $\sqrt{s_{NN}}$ = 200 GeV.
The trigger particles $p_{T}$ range is $3.0<p_{T}<6.0$; the
associate $K_{S}^{0}$, $\Lambda$, or $\overline{\Lambda}$
particles $p_{T}$ range is $1.0<p_{T}<4.0$. The yellow band around
the zero is the systematic errors. } \label{fig1}
\end{figure}

\begin{figure}[htb]
\vspace{-3pt}
\resizebox{0.33\textwidth}{!}{\includegraphics{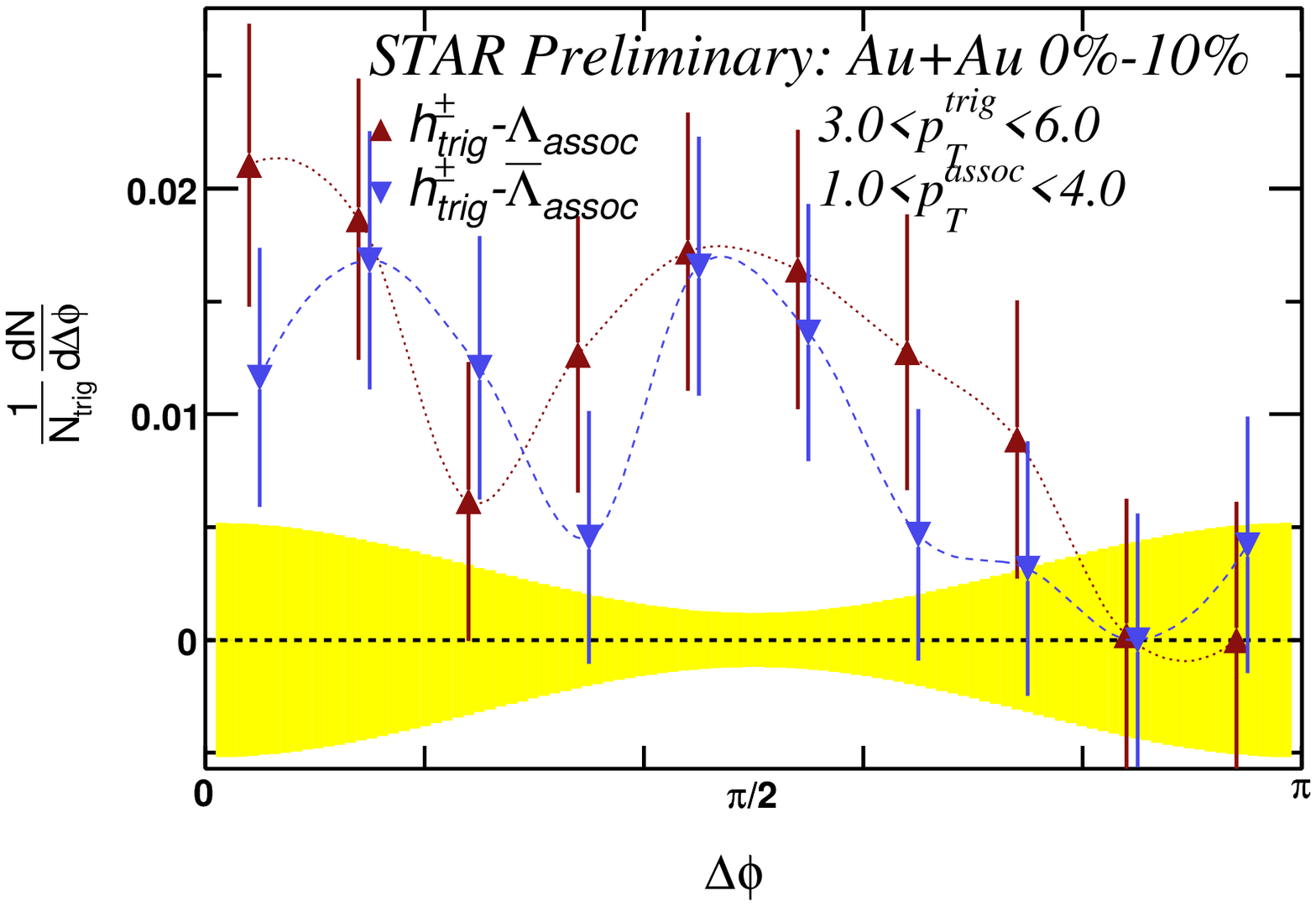}}
\resizebox{0.33\textwidth}{!}{\includegraphics{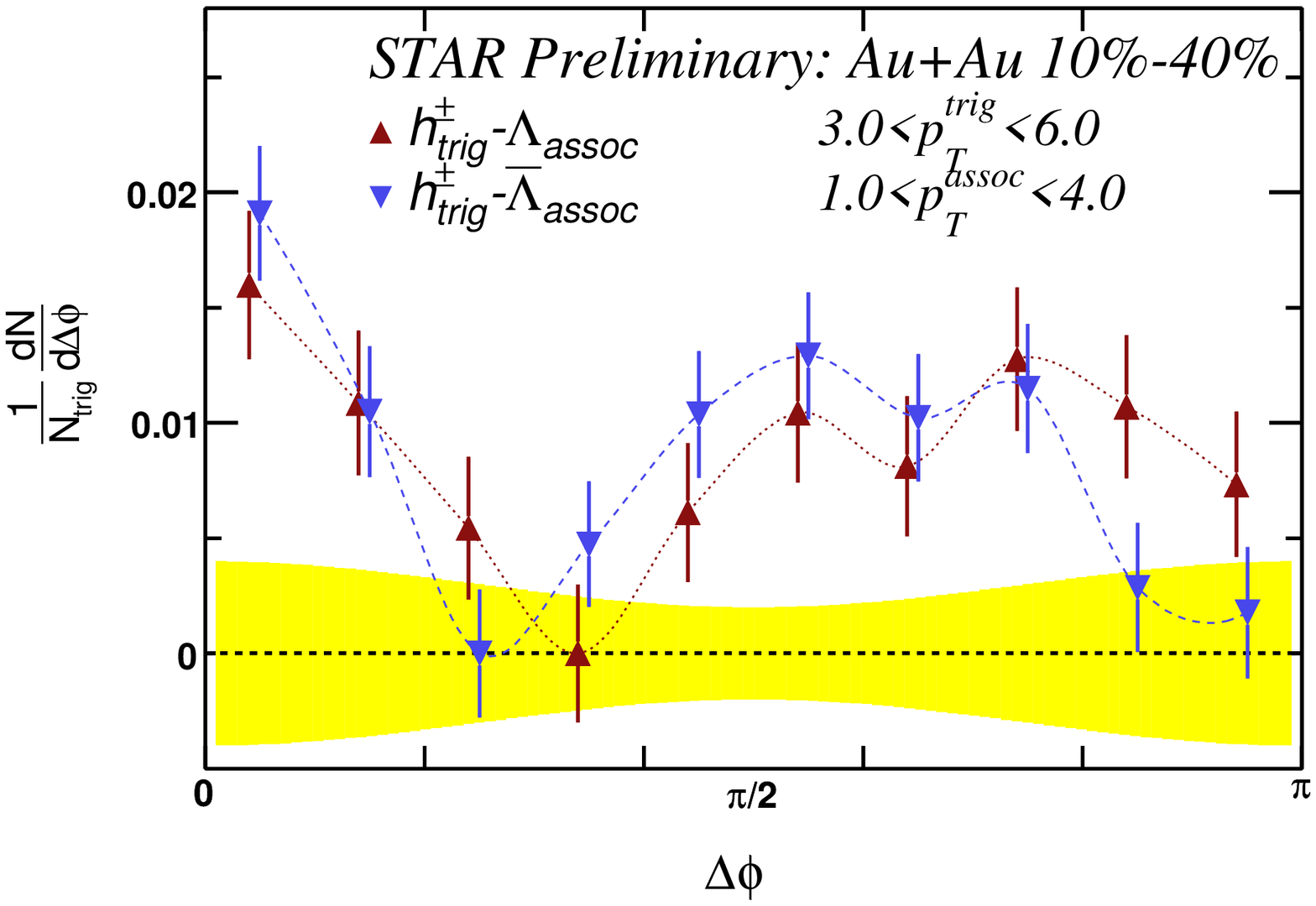}}
\resizebox{0.33\textwidth}{!}{\includegraphics{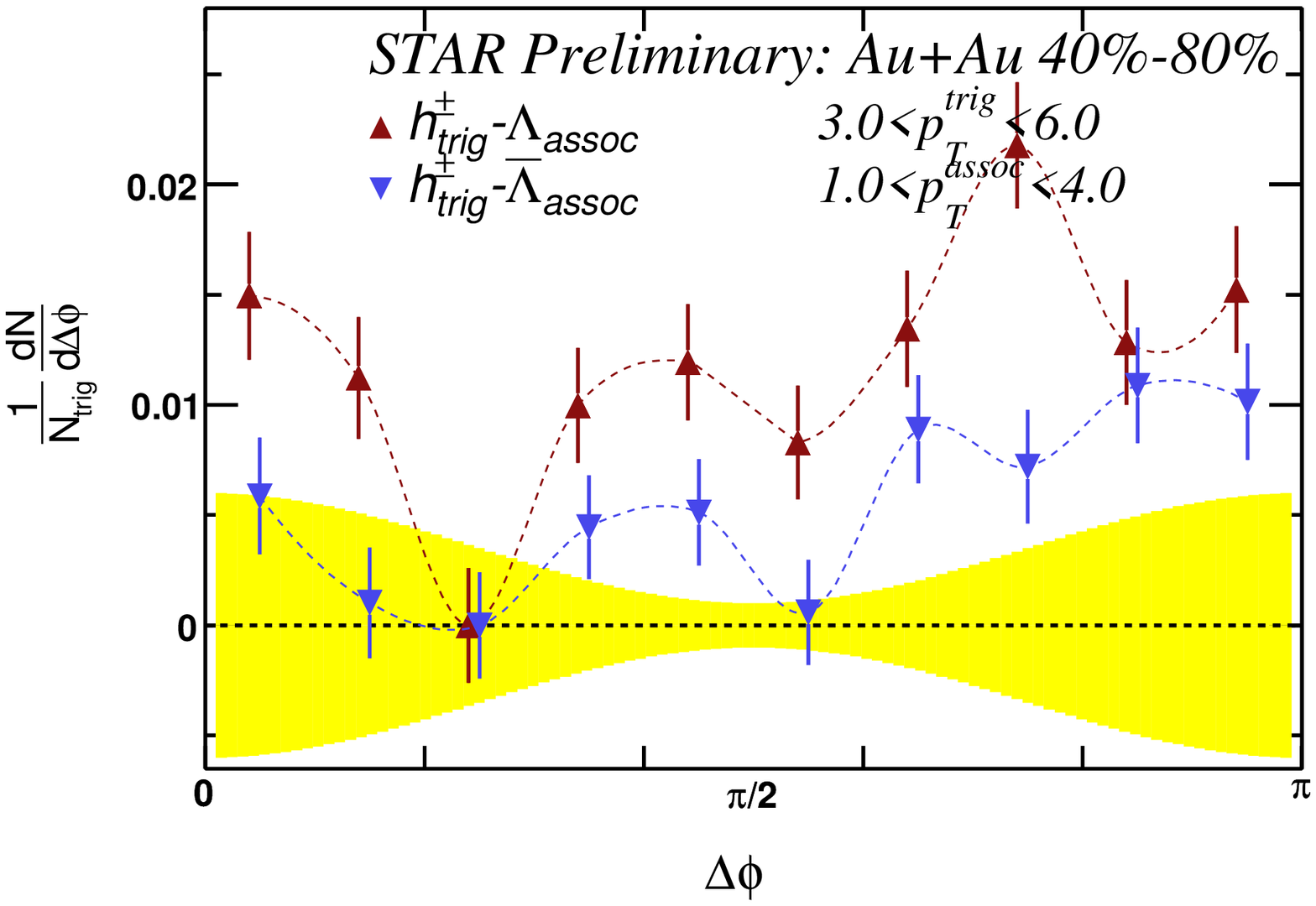}}
\caption[]{Hadron-$\Lambda$, -$\overline{\Lambda}$ correlation
function from central to peripheral collision in $^{197}$Au +
$^{197}$Au collisions at $\sqrt{s_{NN}}$ = 200 GeV. The trigger
particles $p_{T}$ range is $3.0<p_{T}<6.0$; the associate
$\Lambda$, or $\overline{\Lambda}$ particles $p_{T}$ range is
$1.0<p_{T}<4.0$. The yellow band around the zero is the systematic
errors. } \label{fig2}
\end{figure}

The acceptance, efficiency, and background subtracted di-hadron
$dN/d\Delta\phi$ distributions are shown in Fig.~\ref{fig1} and
Fig.~\ref{fig2}. All data are from $\sqrt{s_{NN}}$ = 200 GeV Au+Au
collisions. Fig.~\ref{fig1} shows the hadron-$K_S^0$, and the
hadron-($\Lambda+\overline{\Lambda}$) $dN/d\Delta\phi$
distributions. Fig.~\ref{fig2} shows the hadron-$\Lambda$ and
hadron-$\overline{\Lambda}$ correlations separately. From the left
to right, the plots show the central to peripheral collision. For
all particle combinations a strong correlation is seen on the
near-side of the charged hadron trigger ($\Delta\phi<1.1$) as
would be expected from fragmentation of a fast parton or jet. The
correlation structure on the away-side of the trigger hadron
changes with the collision centrality. From the central to
peripheral, the away-side shows a double bump, broadened and a
single peak. This is consistent with the di-hadron distributions
from STAR~\cite{STAR_Horner}. In the central collision, the
away-sides exhibit a minimum at $\Delta\phi=\pi$ where typically a
maximum would exist. These features are similar to those already
observed for unidentified di-hadron distributions which have much
better statistics~\cite{PHENIX_PRL97_ZYAM}. We extract the
conditional yields of identified $K_{S}^{0}$, $\Lambda$ and
$\overline{\Lambda}$ particles on the near-side
($0.<\Delta\phi<0.35\pi$) and away-side ($0.35\pi<\Delta\phi<\pi$)
of the trigger hadron.

\begin{figure}[htb]
\centering\mbox{ \vspace*{-3pt}
\includegraphics[width=0.70\textwidth]{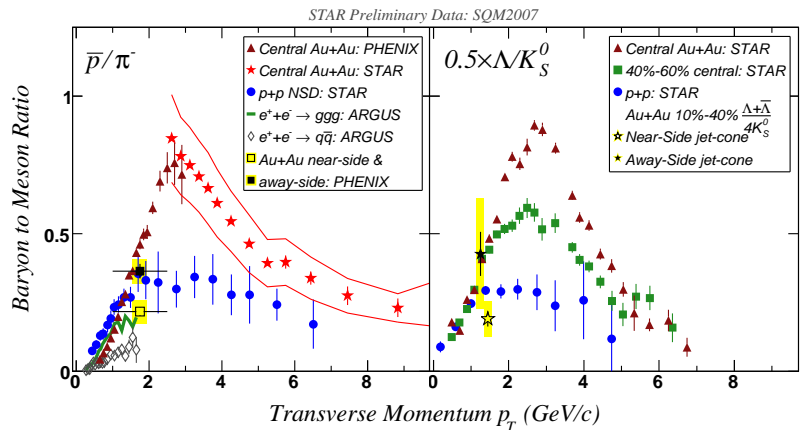}}
\vspace*{-3pt} \caption[]{ Left panel: $\overline{p}$ to $\pi^{-}$
ratio at the mid-rapidity for central Au+Au and p+p collisions at
200 GeV. The measurement of the proton to pion ratio for particles
associated with a trigger hadron ($p_{T}>2.5$) is also shown.
Right panel: $\Lambda$ to $K_{S}^{0}$ ratio in central Au+Au,
mid-peripheral Au+Au and minimum-bias p+p collisions. The
measurement of the $\Lambda$ to $K_{S}^{0}$ ratio for particles
associated with a trigger hadron ($p_{T}>3.0$) is shown. Values
are scaled by 0.5. In the plots, the yellow band is the systematic
errors. } \label{fig3}
\end{figure}

In Fig.~\ref{fig3} we compare our results for the
$(\Lambda+\overline{\Lambda})/K_S^0$ ratio to other measurements
of the baryon-to-meson ratio. The left panel shows the
$\overline{p}/\pi^-$ ratio measured in $e^++e^-$~\cite{argus},
p+p~\cite{STAR_PLB637_Identified_pp}, and
$Au+Au$~\cite{PHENIX_PRL91_ppbar,STAR_PRL97_Identified_AuAu}
collisions (these measurements are not conditional yields). The
right panel shows the $\Lambda/K^0_S$ ratio for p+p,
mid-peripheral $Au+Au$~\cite{STAR_str}, and central $Au+Au$
collisions scaled by 0.5. The measurements of the
$\overline{p}/\pi^-$ ratio made for particles associated with a
trigger hadron ($p_{T}>2.5$) from PHENIX are also shown in the
left panel while our results are shown in the right panel. We find
that both STAR and PHENIX measurements are consistent with a
larger baryon-to-meson ratio on the away-side than on the
near-side. In addition, on the near-side the baryon-to-meson ratio
is closer to values measured in p+p collisions while on the
away-side the ratio is closer to that measured in central or
mid-central Au+Au collisions. This observation may indicate that
the larger parton density of matter is traversed by the away-side
jet. This larger parton density may lead to an enhancement in
baryon production. Such an effect is expected if the baryon
enhancement in the intermediate $p_T$ region observed in Au+Au
collisions is due to multi-parton interactions such as gluon
junction~\cite{gluon_junction} or quark
coalescence~\cite{coalescence}.

More information can be obtained from the distributions in
Fig.~\ref{fig1} and Fig.~\ref{fig2} by examining how the ratios
depend on $\Delta\phi$: \textit{e.g.} the ratios of the
conditional yields on the away-side can help us better understand
the source of the correlations that appear at large angles away
from $\Delta\phi=\pi$. It has been speculated that the enhanced
correlations at wide angles may be related to large angle gluon
radiation~\cite{cherenkov_Koch,Gluonradiation_PLB78_Vitev},
deflection of the away-side jet by the flowing
medium~\cite{Deflect_Armesto}, or a shock wave that is induced in
the medium by a fast moving
parton~\cite{Mach_Casalderrey-Solana,Mach_Stocker}. We expect the
dependence of the particle ratios on $\Delta\phi$ to differ in the
three above scenarios: \textit{e.g.} gluons radiated at large
angles may lead to a larger antibaryon-to-baryon ratio in that
region. A recent study also found that the presence of these
gluons may also lead to an enhanced baryon-to-meson
ratio~\cite{Zhangbu_BaryonDensity}. Alternatively, the higher
density that would be associated with a shock-wave could lead to
an increase in the baryon-to-meson ratio via coalescence of
co-moving partons. It has also been argued that since a shock wave
should be moving at the speed of sound in the medium, the
particles produced from such a shock should not be very fast
particles. For a slow particle to satisfy the $p_T$ cut in our
analysis it would have to be massive. For this reason, one might
expect the correlation at large angles to have a larger number of
massive particles and consequently a larger baryon-to-meson ratio.
Detailed calculations of particle ratios from the above scenarios
have not been made but are being pursued by us.%~\cite{ourNewPaper}.

\begin{figure}[htb]
\vspace{-4pt}
\resizebox{0.45\textwidth}{!}{\includegraphics{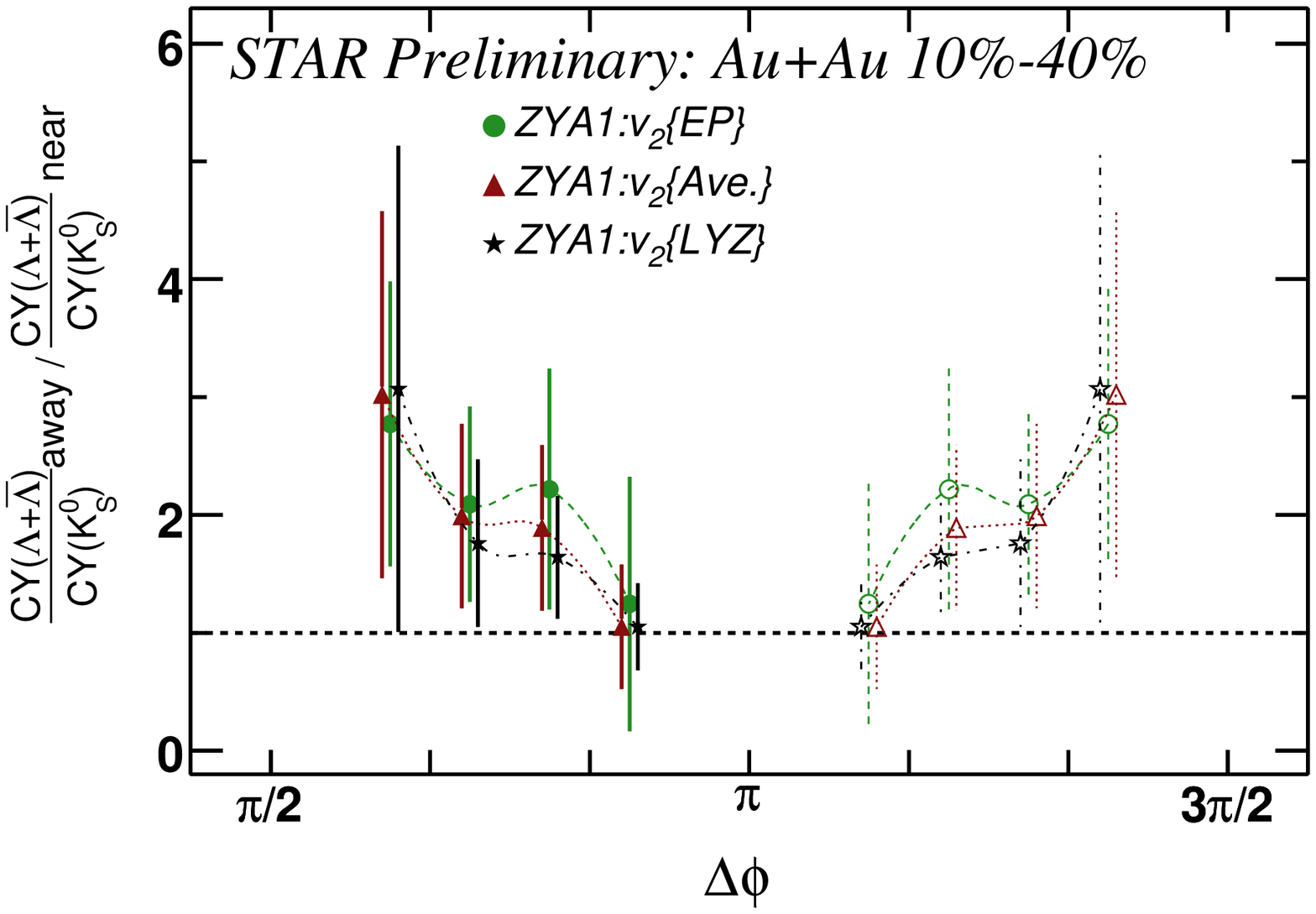}}
\resizebox{0.45\textwidth}{!}{\includegraphics{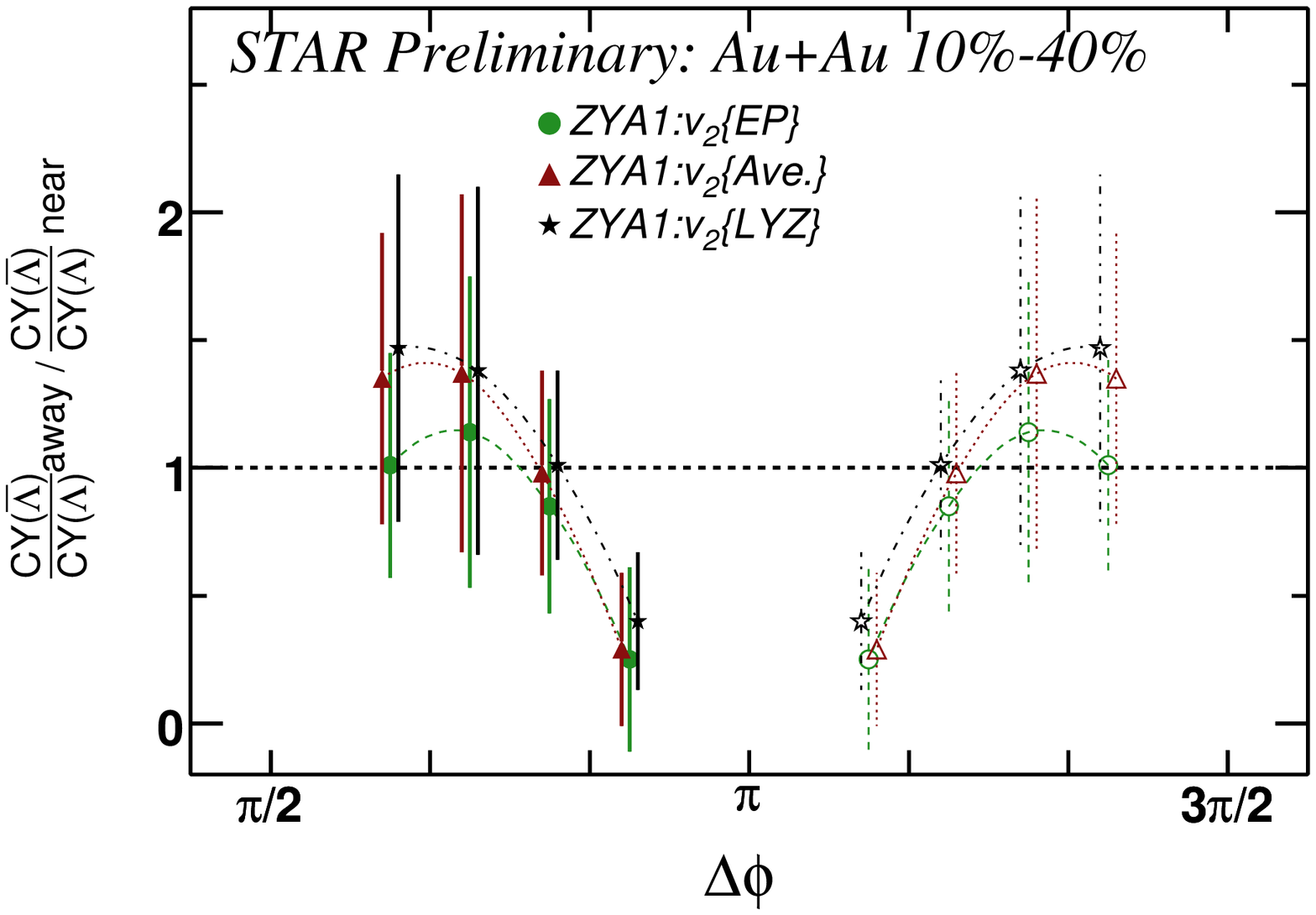}}
\caption[]{Left panel: The baryon-to-meson ratio on the away-side
  vs. $\Delta\phi$ scaled by the same ratio in the near-side
  jet-cone. Data are for $10\%-40\%$ Au+Au collisions at 200 GeV. This
  double ratio appears to be insensitive to the background subtraction
  method. Right panel: the same for $\overline{\Lambda}/\Lambda$.
} \label{fig4}
\end{figure}

Fig.~\ref{fig4} shows the particle ratios (ratios of the
conditional yields) on the away-side as a function of
$\Delta\phi$. The ratios are normalized by the corresponding ratio
measured in the near-side jet-cone so that unity corresponds to
the case where the away-side particle composition is the same as
that in the near-side jet cone. We find that this double ratio is
largely independent of the elliptic flow used in the background
subtraction indicating that such an analysis is able to reduce
systematic uncertainties. The left panel shows the
baryon-to-meson, awayside-to-nearside double ratio and the right
panel shows the antibaryon-to-baryon, awayside-to-nearside double
ratio. In both cases the data from $\Delta\phi<\pi$ (closed
symbols) has been reflected to $\Delta\phi>\pi$ (open symbols).
The uncertainty on both measurements remains large and precludes
strong conclusions about the shape or magnitude of the ratios. We
observe some indication that the
$(\overline{\Lambda}+\Lambda)/K_S^0$ ratios may be large at around
$\Delta\phi=\pi/2$ than they are at $\Delta\phi=\pi$. This may be
consistent, for example, with the increase of the parton density
at large angles as discussed above.

\section{Summary}\label{sum}
We measured di-hadron azimuthal angle correlations in Au+Au
collisions at $\sqrt{s_{NN}}$ = 200 GeV. Charged hadrons
($3.0<p_{T}<6.0$ GeV/c) are used as the trigger particle;
$K_{S}^{0}$s, $\Lambda$s and $\overline{\Lambda}$s
($1.0<p_{T}<4.0$ GeV/c) are used as the associated particles. We
extracted the conditional yields of identified associate particles
on the near- and away-side of the jet trigger and calculated the
near and away-side particle ratios.  The systematic uncertainty
from $v_{2}$ and the background normalization are large. These
uncertainties can be reduced with more data to reduce the error on
the level of the background and a better understanding of $v_{2}$
to reduce uncertainty on the shape of the background. Both STAR
and PHENIX results are consistent with a larger baryon-to-meson
ratio on the away-side than the near-side. We studied the shape of
away-side particle ratios and find that this shape is insensitive
to several sources of systematic uncertainty. Our measurements are
consistent with the physical picture that the parton density may
be higher at large angles away from $\Delta\phi=\pi$. These
measurements should help elucidate how fast partons interact with
the matter created in Au+Au collisions at RHIC.

This work is completed with P. Sorensen. We thank H. Huang, Z. Xu,
A. H. Tang and the STAR Group at BNL for enlightening discussions.
This work was supported in part by the National Natural Science
Foundation of China under Grant No. 10610285, the Knowledge
Innovation Project of Chinese Academy of Sciences under Grant No.
KJCX2-YW-A14 and KJXC3-SYW-N2, and the Shanghai Development
Foundation for Science and Technology under Grant Numbers
05XD14021 and "Flying Plan" of Shanghai Association for Science
and Technology.

\vfill\eject

\section*{References}

\begin{thebibliography}{34}

\bibitem{STAR_PRL90_flow}
  C.~Adler {\it et al.}  [STAR Collab.],
  %``Azimuthal anisotropy and correlations in the hard scattering regime at
  %RHIC,''
  {\it Phys.\ Rev.\ Lett.\  } {\bf 90}, 032301 (2003).

\bibitem{STAR_PRL90_flow2}
  J.~Adams {\it et al.}  [STAR Collab.],
  %``Azimuthal Anisotropy in Au+Au Collisions at sqrt(sNN) = 200 GeV''
  {\it Phys.\ Rev.\ } C {\bf 72}, 0140904 (2005).

\bibitem{STAR_NPA_whitepapers}
  J.~Adams {\it et al.}  [STAR Collab.],
  %``Experimental and theoretical challenges in the search for the quark  gluon
  %plasma: The STAR collaboration's critical assessment of the  evidence from
  %RHIC collisions,''
  {\it Nucl.\ Phys.\ } A {\bf 757}, 102 (2005).

\bibitem{STAR_PRL90_Hardtke}
  C.~Adler {\it et al.}  [STAR Collab.],
  %``Disappearance of back-to-back high p(T) hadron correlations in central Au +
  %Au collisions at s(NN)**(1/2) = 200-GeV,''
  {\it Phys.\ Rev.\ Lett.\  } {\bf 90}, 082302 (2003).

\bibitem{STAR_PRC73_2006}
   J.~Adams {\it et al.}  [STAR Collab.],
   %``Minijet deformation and charge-independent angular correlations
   %on momentum
   %subspace (eta, phi) in Au-Au collisions at s(NN)**(1/2) = 130-GeV,''
   {\it Phys.\ Rev.\ } C {\bf 73}, 064907 (2006).

\bibitem{STAR_PRC75_2007}
   J.~Adams {\it et al.}  [STAR Collab.],
   %``Delta(phi) Delta(eta) correlations in central Au + Au
   %collisions at
   %s(NN)**(1/2) = 200-GeV,''
   {\it Phys.\ Rev.\ } C {\bf 75}, 034901 (2007).

\bibitem{STAR_PRL91_highpt}
  J.~Adams {\it et al.}  [STAR Collab.],
  %``Evidence from d + Au measurements for final-state suppression of high  p(T)
  %hadrons in Au + Au collisions at RHIC,''
  {\it Phys.\ Rev.\ Lett.\  } {\bf 91}, 072304 (2003).

\bibitem{STAR_PRL95_Fuqiang}
  J.~Adams {\it et al.}  [STAR Collab.],
  %``Distributions of charged hadrons associated with high transverse  momentum
  %particles in p p and Au + Au collisions at s(NN)**(1/2) =  200-GeV,''
  {\it Phys.\ Rev.\ Lett.\  } {\bf 95}, 152301 (2005).

\bibitem{volcano}
  M.~McCumber and J.~Frantz,
  %``Return of the volcano: PHENIX azimuthal correlations 62.4-GeV Au + Au,''
  {\it Acta Phys.\ Hung.\ } A  {\bf 27}, 213 (2006).

\bibitem{Mach_Casalderrey-Solana}
  J.~Casalderrey-Solana,
  %``Mach cones in quark gluon plasma,''
  {\it arXiv:hep-ph/0701257};
%\bibitem{Casalderrey-Solana:2004qm}
   %J.~Casalderrey-Solana {\it et al.},
   %``Conical flow induced by quenched QCD jets,''
   {\it Nucl.\ Phys.\ } A  {\bf 774}, 577 (2006).

\bibitem{Mach_Stocker}
  H.~Stocker, B.~Betz and P.~Rau,
  %``Hydrodynamic flow and jet induced mach shocks at RHIC and LHC,''
  {\it arXiv:nucl-th/0703054}.

\bibitem{Deflect_Armesto}
  N.~Armesto, C.~A.~Salgado and U.~A.~Wiedemann,
  %``Low-p(T) collective flow induces high-p(T) jet quenching,''
  {\it Phys.\ Rev.\ } C {\bf 72}, 064910 (2005).

\bibitem{cherenkov_Koch}
  V.~Koch, A.~Majumder and X.~N.~Wang,
  %``Cherenkov radiation from jets in heavy-ion collisions,''
  {\it Phys.\ Rev.\ Lett.\  } {\bf 96}, 172302 (2006).

\bibitem{MaGL_AMPT}
  G.~L.~Ma {\it et al.},
  %``Di-hadron azimuthal correlation and Mach-like cone structure in parton  /
  %hadron transport model,''
  {\it Phys.\ Lett.\ } B {\bf 641}, 362 (2006)

%  G.~L.~Ma {\it et al.},
%  %``Three-particle correlations from strong partonic cascade in the AMPT
%  %model,''
%  Phys.\ Lett.\  B {\bf 647}, 122 (2007)

\bibitem{tri-hadron_Ulery}
  J.~G.~Ulery  [STAR Collab.],
  %``Three-particle azimuthal correlations,''
  {\it Nucl.\ Phys.\ } A {\bf 783}, 511 (2007);

%  J.~G.~Ulery  [STAR Collab.],
%  %``Are There Mach Cones in Heavy Ion Collisions? Three-Particle   Correlations
%  %from STAR,''
%  arXiv:0704.0224 [nucl-ex].

\bibitem{tri-hadron_Ajitanand_PHENIX}
  N.~N.~Ajitanand  [PHENIX Collab.],
  %``Two and three particle flavor dependent correlations,''
  {\it Nucl.\ Phys.\ } A {\bf 774}, 585 (2006);

  %N.~N.~Ajitanand  [PHENIX Collab.],
  %``Three Particle Correlation Functions: A Probe For Mach Cones At Rhic,''
  %AIP Conf.\ Proc.\  {\bf 842}, 122 (2006).

\bibitem{P.Sorensen}
  P.~R.~Sorensen,
  %``Identified particle spectra and jet interactions with the medium,''
  {\it Nucl.\ Phys.\ } A {\bf 774}, 247 (2006)

\bibitem{PHENIX_PRL91_ppbar}
  S.~S.~Adler {\it et al.}  [PHENIX Collab.],
  %``Scaling properties of proton and anti-proton production in s(NN)**(1/2)  =
  %200-GeV Au + Au collisions,''
  {\it Phys.\ Rev.\ Lett.\  } {\bf 91}, 172301 (2003).

\bibitem{quarkvsgluon}
  B.~I.~Abelev {\it et al.}  [STAR Collab.],
  %``Strange particle production in p + p collisions at s**(1/2) = 200-GeV,''
  {\it arXiv:nucl-ex/0607033}.

\bibitem{Zhangbu_BaryonDensity}
  H.~d.~Liu and Z.~Xu,
  %``Universal anti-baryon density in e+ e-, gamma p, p p, p A and A A
  %collisions,''
  {\it arXiv:nucl-ex/0610035}.

%\bibitem{STAR_TPC}
%  K.~H.~Ackermann {\it et al.}  [STAR Collab.],
%  %``STAR detector overview,''
%  Nucl.\ Instrum.\ Meth.\  A {\bf 499}, 624 (2003).

\bibitem{STAR_Horner}
  M.~J.~Horner  [STAR Collab.],
  %``Low- and intermediate-p(T) di-hadron distributions in Au + Au collisions at
  %s(NN)**(1/2) = 200-GeV from STAR,''
  %{\it arXiv:nucl-ex/0701069}.
  {\it J Phys G: Nucl.\ Part.\ Phys.} {\bf 34}, S995 (2007).

\bibitem{PHENIX_PRL97_ZYAM}
   S.~S.~Adler {\it et al.}  [PHENIX Collab.],
   %``Modifications to di-jet hadron pair correlations in Au + Au
   %collisions  at
   %s(NN)**(1/2) = 200-GeV,''
   {\it Phys.\ Rev.\ Lett.\  } {\bf 97}, 052301 (2006).

\bibitem{STAR_PRC66_v24}
  C.~Adler {\it et al.}  [STAR Collab.],
  %``Elliptic flow from two- and four-particle correlations in Au + Au
  %collisions at s(NN)**(1/2) = 130-GeV,''
  {\it Phys.\ Rev.\ } C {\bf 66}, 034904 (2002).

\bibitem{LYZ}
  R.~Bhalerao {\it et al.},
  %``Analysis of anisotropic flow with Lee-Yang zeroes,''
  {\it Nucl.\ Phys.\ } A {\bf 727}, 373 (2003).

\bibitem{STAR_v2fluctuation}
  P.~Sorensen  [STAR Collab.],
  %``Elliptic flow fluctuations in Au + Au collisions at s(NN)**(1/2) =
  %200-GeV,''
  {\it arXiv:nucl-ex/0612021}.

\bibitem{STAR_PRL89_KsLv2}
  C.~Adler {\it et al.}  [STAR Collab.],
  %``Azimuthal anisotropy of K0(S) and Lambda + anti-Lambda production at
  %mid-rapidity from Au + Au collisions at s(NN)**(1/2) = 130-GeV,''
  {\it Phys.\ Rev.\ Lett.\  } {\bf 89}, 132301 (2002).

\bibitem{STAR_PRL92_partdepend}
  J.~Adams {\it et al.}  [STAR Collab.],
  %``Particle dependence of azimuthal anisotropy and nuclear modification of
  %particle production at moderate p(T) in Au + Au collisions at  s(NN)**(1/2) =
  %200-GeV,''
  {\it Phys.\ Rev.\ Lett.\  } {\bf 92}, 052302 (2004).

\bibitem{STAR_PRL95_Flow_KL}
  J.~Adams {\it et al.}  [STAR Collab.],
  %``Multi-strange baryon elliptic flow in Au + Au collisions at  s(NN)**(1/2) =
  %200-GeV,''
  {\it Phys.\ Rev.\ Lett.\  } {\bf 95}, 122301 (2005).

\bibitem{argus}
  H.~Albrecht {\it et al.}  [ARGUS Collab.],
  %``INCLUSIVE PRODUCTION OF CHARGED PIONS, CHARGED AND NEUTRAL KAONS AND
  %ANTI-PROTONS IN e+ e- ANNIHILATION AT 10-GeV AND IN DIRECT UPSILON DECAYS,''
  {\it Z.\ Phys.\ } C {\bf 44}, 547 (1989).

\bibitem{STAR_PLB637_Identified_pp}
  J.~Adams {\it et al.}  [STAR Collab.],
  %``Identified hadron spectra at large transverse momentum in p + p and d +  Au
  %collisions at s(NN)**(1/2) = 200-GeV,''
  {\it Phys.\ Lett.\ } B {\bf 637}, 161 (2006).

\bibitem{STAR_PRL97_Identified_AuAu}
  B.~I.~Abelev {\it et al.}  [STAR Collab.],
  %``Identified baryon and meson distributions at large transverse momenta  from
  %Au + Au collisions at s(NN)**(1/2) = 200-GeV,''
  {\it Phys.\ Rev.\ Lett.\  } {\bf 97}, 152301 (2006).

\bibitem{STAR_str}
  J.~Adams {\it et al.}  [STAR Collab.],
  %``Measurements of identified particles at intermediate transverse  momentum
  %in the STAR experiment from Au + Au collisions at s(NN)**(1/2)  = 200-GeV,''
  {\it arXiv:nucl-ex/0601042}.

\bibitem{gluon_junction}
  D.~Kharzeev,
  %``Can Gluons Trace Baryon Number?,''
  {\it Phys.\ Lett.\ } B {\bf 378}, 238 (1996);
%\bibitem{Vance:1998vh}
  S.~E.~Vance, M.~Gyulassy and X.~N.~Wang,
  %``Baryon number transport via gluonic junctions,''
  {\it Phys.\ Lett.\ } B {\bf 443}, 45 (1998);
%\bibitem{Vitev:2002wh}
  I.~Vitev and M.~Gyulassy,
  %``High-p(T) pion quenching versus anti+baryon enhancement in nucleus nucleus
  %collisions,''
  {\it Nucl.\ Phys.\ } A {\bf 715}, 779 (2003).

\bibitem{coalescence}
  D.~Molnar {\it et al.},
  %``Elliptic flow at large transverse momenta from quark coalescence,''
  {\it Phys.\ Rev.\ Lett.\  } {\bf 91}, 092301 (2003);
%\bibitem{Hwa:2003bn}
  R.~C.~Hwa {\it et al.},
  % ``Scaling distributions of quarks, mesons and proton for all p(T), energy
  %and centrality,''
  {\it Phys.\ Rev.\ } C {\bf 67}, 064902 (2003);
%\bibitem{Fries:2003kq}
  R.~J.~Fries, B.~Muller, {\it et al.},
  % ``Hadron production in heavy ion collisions: Fragmentation and  recombination
  %from a dense parton phase,''
  {\it Phys.\ Rev.\ } C {\bf 68}, 044902 (2003);
%\bibitem{Greco:2003mm}
  V.~Greco, {\it et al.},
  %``Parton coalescence at RHIC,''
  {\it Phys.\ Rev.\ } C {\bf 68}, 034904 (2003).

\bibitem{Gluonradiation_PLB78_Vitev}
   I.~Vitev,
   %``Large angle hadron correlations from medium-induced gluon
   %radiation,''
   {\it Phys.\ Lett.\ } B {\bf 630}, 78 (2005).

%\bibitem{ourNewPaper} J.X. Zuo and P. Sorensen, {\it In preparation}.

\end{thebibliography}
\end{document}